# Observation of exceptional topology and nonlocal skin effect in Klein bottle electric circuits

Pengtao Lai[1], Xiangru Chen[2], Yugan Tang[1], Yejian Hu[2], Zhenhang Pu[2], Hui Liu[1], Weiyin Deng[2*], Hua Cheng[1*], Zhengyou Liu[2,3], and Shuqi Chen[1,4,5*]

[1] The Key Laboratory of Weak Light Nonlinear Photonics, Ministry of Education, School of Physics and TEDA Institute of Applied Physics, Nankai University, Tianjin 300071, China
[2] Key Laboratory of Artificial Micro- and Nanostructures of Ministry of Education and School of Physics and Technology, Wuhan University, Wuhan 430072, China
[3] Institute for Advanced Studies, Wuhan University, Wuhan 430072, China
[4] School of Materials Science and Engineering, Smart Sensing Interdisciplinary Science Center, Nankai University, Tianjin 300350, China
[5] The Collaborative Innovation Center of Extreme Optics, Shanxi University, Taiyuan, Shanxi 030006, China

*Corresponding author.
Emails: dengwy@whu.edu.cn; hcheng@nankai.edu.cn; schen@nankai.edu.cn

Symmetry and its representation play a crucial role in topological phases, including both Hermitian and non-Hermitian paradigms. In the presence of synthetic gauge field, spatial symmetries should be projectively represented, which can modify the Brillouin manifold. However, this is often overlooked in non-Hermitian systems. Here, we present that momentum-space non-symmorphic reflection symmetry, a typical projective symmetry, induce exceptional topology and the nonlocal skin effect in a two-dimensional non-Hermitian electric circuit. We observe the total topological charges 2, rather than 0, for all exceptional points in a Brillouin Klein bottle manifold, and the phase transition when an exceptional point crosses the antiparallel boundary and flips its topological charge. We further observe a novel skin effect that the skin modes at one side are nonlocally connected to those on the opposite side separated by half of the reciprocal lattice. Our results unveil the unique non-Hermitian phenomena enabled by the projective symmetry, and open avenues for exploring the non-Hermitian topology beyond Brillouin torus manifold.

Over the past decades, symmetry has provided a theoretical framework for the classification of topological matter [1,2]. In the presence of gauge field, an additional phase factor is introduced to correct the group multiplication relation, giving rise to the so-called projective symmetry [3-10]. A paradigmatic example is that the $\pi$ gauge flux alters the commutation relation between reflection symmetry and translation symmetry, transforming it into an anti-commutation relation, which forms the momentum-space non-symmorphic (MSNS) reflection symmetry [11-19]. Under this symmetry, the Brillouin torus can be reduced to non-orientable manifolds [20], such as the Brillouin Klein bottle (BKB) [12] and the Brillouin real projective plane [16]. As the manifold of the Brillouin zone (BZ) is modified, a series of novel topological insulator phases emerge [12,13,16-18], such as the Klein bottle insulator, which features two nonlocally related edge states [12,13]. Furthermore, for topological semimetals, recent studies have demonstrated that in non-orientable manifolds, the total chirality of Weyl points can be nonzero, thereby circumvent the constraints imposed by the well-known "no-go" theorem of Weyl points [21].

Recently, non-Hermitian systems featured with complex spectra exhibit novel topological phenomena such as exceptional points (EPs) and skin effects, in which symmetry also stands as a cornerstone [22-24]. For EPs, not only do eigenvalues become degenerate, but the corresponding eigenvectors also coalesce into a single vector [25-29]. EPs carry nontrivial topological charges characterized by the discriminant number [30-33]. A key corollary is that EPs with opposite topological charges must appear in pairs within the BZ, which is the fermion doubling theorem of EPs [30]. This theorem relies on the assumption that the BZ manifold is a torus and may not hold if the manifold is altered [34-36]. For skin effect, a macroscopic number of bulk states are localized at the boundary [37-39]. To date, various skin effects with distinct real-space configurations have been identified, including the geometry-dependent skin effect [40-44], higher-order skin effect [45-48], and hybrid skin effect [49-52]. When projective symmetry is taken into consideration, more intriguing phenomena arise in non-Hermitian systems [53-56], such as the $Z_2$ skin effect [55]. However, the study of projective symmetry and ensuing novel Brillouin manifold in non-Hermitian systems remains in its infancy, and their roles in exceptional topology and skin effects remain insufficiently explored.

In this work, we realize a non-Hermitian Klein bottle electric circuit, and reveal the exceptional topology and nonlocal skin effect induced by MSNS reflection symmetry within the system. The circuit platform has distinct advantages for introducing non-Hermiticity and measuring complex spectra [57-59]. We first illustrate the intriguing non-Hermitian phenomena in the lattice model. In the circuit experiments, we then observe that the total topological charge of all EPs in the BKB is 2, which circumvent the famous Fermion-doubling theorem. And we further demonstrate a novel phase transition of EPs, where an EP crosses the antiparallel boundary of BKB and flips its topological charge. Finally, we reveal a nonlocal skin effect that the skin modes at one side are nonlocally connected to those on the opposite side separated by half of the reciprocal lattice, protected by MSNS reflection symmetry. The experimental results agree well with the theoretical predictions, confirming the exceptional topology in the BKB and the nonlocal skin effect.

We start from the tight-binding model shown in Fig. 1(a), where each unit cell consists of two sublattices. The red and blue tubes represent the negative and positive intercell couplings, denoted as $\pm t_3 e^{i\theta}$, respectively. The reciprocal intercell coupling along the $x$-direction is denoted by $t_1$, and non-Hermitian terms are introduced through the loss term $i\gamma$ in the intracell coupling $t_2 + i\gamma$. The Hamiltonian in momentum space is given by

$$H(k_x, k_y) = (t_1 \cos k_x + t_2 + i\gamma)\sigma_x + [t_1 \sin k_x + 2t_3 \sin(k_y - \theta)]\sigma_y, \quad (1)$$

where $\sigma_x$ and $\sigma_y$ are Pauli matrices. The non-Hermitian Hamiltonian $H(k_x, k_y)$ exhibits the MSNS reflection symmetry $M_x^g = \sigma_x$ and satisfies $M_x^g H(k_x, k_y)(M_x^g)^{-1} = H(-k_x, k_y + \pi)$. Under this symmetry, the eigenvalues and eigenvectors satisfy $H(-k_x, k_y + \pi) M_x^g \varphi(k_x, k_y) = E(k_x, k_y) M_x^g \varphi(k_x, k_y)$, indicating the band topology of the lattice can be fully captured within half of the BZ. Figure 1(b) illustrates the evolution of the energy as a function of $k_x$ at $k_y = 5\pi/6$ and $k_y = 11\pi/6$. As the spectrum at $(k_x, k_y)$ is equivalent to that at $(-k_x, k_y + \pi)$, the two band dispersions wind in opposite directions along $k_x$. For convenience in characterizing the phase transition and annihilation processes of EPs, we select the region $(k_x, k_y) \in [0, 2\pi] \times \left[\frac{5\pi}{6}, \frac{11\pi}{6}\right]$ as the reduced BZ. Notably, the upper and lower boundaries of the reduced BZ are oppositely oriented as they are related by the MSNS reflection symmetry $M_x^g$. Consequently, the reduced BZ forms a BKB in momentum space.

In the BKB, EPs exhibit a series of topological properties absent in conventional BZ. Figure 1(c) shows the phase of $\det[H(k_x, k_y)]$ in the BKB. The topology of the EPs can be characterized by the discriminant number $v(k_D) = \frac{1}{2\pi} \oint_\Gamma dk \, \ln\det[H(k) - E(k_{\text{EP}})]$, where $\Gamma$ is the loop counterclockwise encircling the EP and $E(k_{\text{EP}}) = 0$ is the eigenvalue of EP [37]. Since the variation of the modulus of $\det[H(k) - E(k_{\text{EP}})]$ is trivial, the topology of the EPs can be revealed by the phase of $\det[H(k) - E(k_{\text{EP}})]$. There are two EPs located at $(k_x, k_y) = (\pm\pi/2, \pi)$, corresponding to the vortices in the phase diagram, and the topological charge of each EP can be inferred from the direction of rotation of these phase vortices. It is observed that the two EPs in the BKB have a topological charge of $+1$, which violates the fermion doubling theorem. In fact, in the BKB, the sum of the topological charge of all EPs must belong to $2\mathbb{Z}$. Moreover, as the top and bottom boundaries of the BKB are oppositely oriented, when an EP crosses the top and bottom boundaries of BKB, its topological charge flips sign. We present the first concrete model to demonstrate this novel phase transition. As $t_2$ increases from 0 to 1, two EPs at $k_y = \pi$ move in opposite directions, as shown in the upper panel of Fig. 1(d). When one of these EPs crosses the lower boundary, an EP with opposite topological charge enters the upper boundary. The lower panel of Fig. 1(d) shows the evolution of the total topological charge, $v_{\text{EP}}$, of the EPs in the BKB as a function of $t_2$. As an EP undergoes a phase transition upon crossing the boundary of the BKB, the total topological charge in the BKB changes from 2 to 0. Further increasing $t_2$ results in the annihilation of the two EPs at $t_2 = 1$, opening a band gap
4

(Supplementary Material [60]). This indicates that, in the BKB, although the EPs have the same topological charge, they can still annihilate each other by crossing the boundary of the BKB.

The existence of EPs guarantees the emergence of skin effect, while symmetries impose constraints on the skin effect. The skin effect along the direction $n$ can be characterized by the spectral winding number $v_n = \frac{1}{2\pi i}\oint_L dk \nabla_k \ln\det[H(k) - E_0]$, where $L$ denotes the path of integration in the BZ along the $k_n$-direction, and $E_0$ is the reference energy. The skin effect appears at the boundaries in the direction where the spectral winding number $v_n \neq 0$, and is absent at boundaries where $v_n = 0$. At $\theta = 0$, besides the MSNS reflection symmetry, the Hamiltonian $H(k_x, k_y)$ also respects inversion symmetry $P = \sigma_x$ and anomalous time reversal symmetry $T$, which satisfy $PH(k_x, k_y)(P)^{-1} = H(-k_x, -k_y)$, and $H(k_x, k_y) = H(-k_x, -k_y)^T$, respectively. Since $\det[H] = \det[H^T]$, both $P$ and $T$ impose identical constraints on the spectral winding number. Therefore, we take $M_x^g$ and $P$ as examples to discuss the constraints on spectral winding numbers in the spinless system (Supplementary Material [60]):

$$v_x(k_y) \stackrel{M_x^g}{=\!=} -v_x(k_y + \pi), \tag{2a}$$

$$v_y(k_x) \stackrel{M_x^g}{=\!=} v_y(-k_x), \tag{2b}$$

$$v_x(k_y) \stackrel{P}{=\!=} -v_x(-k_y), \tag{2c}$$

$$v_y(k_x) \stackrel{P}{=\!=} -v_y(-k_x). \tag{2d}$$

The analysis from symmetry reveals two key features of the skin effect. For one thing, with regard to the skin effect along the $x$-direction, Eq. (2a) indicates that the MSNS reflection symmetry nonlocally maps the skin modes at $k_y$ to those on the opposite side at $k_y + \pi$. Similarly, Eq. (2c) implies that skin modes at $k_y$ and $-k_y$ are localized at opposite boundaries. Figure 1(e) shows the distribution of all eigenstates with respect to $k_y$, $W(k,j) = \frac{1}{N}\sum_n |\psi_n(k,j)|^2$, where $\psi_n(k,j)$ is the $n$th normalized right eigenstate at site $j$, $k$ is the momentum, and $N$ is the total number

of eigenstates. The eigenstates with $k_y \in (0, \pi)$ are localized at the boundary opposite to those with $k_y \in (\pi, 2\pi)$, which selectively filters the eigenmodes with $k_y < 0$ and $k_y > 0$. At $k_y = 0$ and $k_y = \pi$, eigenstates exhibit Bloch-wave-like extended modes, serving as transitional states between the left and right skin effects. For another, for the skin effect along the $x$-direction, combining Eqs. (2b) and (2d) yields $v_y(k_x) = 0$, indicating the absence of skin effect at the boundaries in the $y$-direction. Figure 1(f) shows the distribution of eigenstates in the $y$-direction, where no skin effect appears, and all eigenstates are Bloch-wave-like extended modes. The spatial distributions of the skin effect in Figs. 1(e) and 1(f) are consistent with the result derived from the spatial symmetry in Eq. (2).

It is worth noting that the skin effect can be modulated by $\theta$. When $\theta \neq 0$ or $\pi$, both inversion symmetry $P$ and anomalous time reversal symmetry $T$ are broken. As a result, the skin modes at $k_y$ and $-k_y$ are no longer required to be localized at opposite boundaries. Furthermore, by adjusting $\theta$, the skin effect localized at the left boundary shifts from $[0, \pi]$ to $[\theta, \pi + \theta]$, effectively sliding the filter window (Supplementary Material [60]).

We implement a passive electric circuit to realize the tight-binding model with $\theta = 0$. Figure 2(a) displays the fabricated electric circuit sample with $12 \times 12$ unit cells, and the corresponding schematic is shown in Fig. 2(b). Each unit consists of two sites, which are grounded through an $LC$ resonator ($L$ and $C_4$). Using Kirchhoff's law, $I = JV$, where $I$ and $V$ denote the input current and voltage at each node, the circuit Laplacian (or admittance matrix) $J$ can be expressed as

$$\frac{J}{i\omega} = \left(-C_1 \cos k_x - C_2 + \frac{i}{\omega R} + t_y\right) \sigma_x - (C_1 \sin k_x + t_{y1})\sigma_y + \varepsilon_0, \quad (3)$$

where $t_y = \left(\frac{1}{\omega^2 L} - C_3\right) \cos k_y$, and $t_{y1} = (C_3 + \frac{1}{\omega^2 L}) \sin k_y$. $\varepsilon_0 = C_1 + C_2 + C_3 + C_4 - \frac{1}{\omega^2 L} - \frac{i}{\omega R}$ is a constant term for a fixed angular frequency $\omega = 2\pi f$, which does not affect the topology of the system. The correspondence between the tight-binding model and the electric circuit is as follows: the lossy coupling $i\gamma$ is introduced by the resistor $R$; the real intercell and intracell couplings, $t_1$ and $t_2$, are realized through capacitors $C_1$ and $C_2$; the positive and negative couplings are constructed through the capacitors $C_3$ and inductors $\frac{1}{\omega^2 L}$. At the critical angular frequency $\omega = \frac{1}{\sqrt{LC_3}}$, the positive and negative couplings in the circuit become equal, and the circuit Laplacian $J$ takes the same form as the Hamiltonian in Eq. (1). In the experiment, the admittance matrix $J$ can be obtained by measuring the voltage and current response at each node of the circuit (Supplementary Material [60]).

To reveal the exceptional topology in the BKB, we measure the admittance spectrum around each exceptional point. The periodic boundary conditions are constructed by connect the left and right, as well as the top and bottom boundaries of the sample. Figure 3(a) shows the periodic admittance spectrum when the intracell capacitor $C_2$ is not connected, corresponding to the case where $t_2 = 0$ in the tight-binding model. At $k_y = \pi$, there are two EPs with a topological charge of $+1$. In the experiment, we measure the braiding of two bands around each EP along loop 1 and loop 2 via a stroboscopic approach. As shown in Figs. 3(b) and 3(c), the two energy bands wind in the same direction (counterclockwise), corresponding to the two EPs with the same topological charge. Therefore, in the BKB, the sum of the topological charges of all EPs is not zero but belongs to $2\mathbb{Z}$. Additionally, when the 18 μH capacitor $C_2$ is connected to the circuit, one of the EPs crosses the boundary of the BKB and undergoes a sign flip of its topological charge, as shown in Fig. 3(d). Figures 3(e) and 3(f) show the stroboscopic evolution of energy bands along loop 3 and loop 4, respectively. In Fig. 3(e), the two bands wind counterclockwise, while they wind clockwise in Fig. 3(f), corresponding to the two EPs having opposite topological charges. This further confirms that the topological phase transition occurs when the EPs cross the antiparallel boundary of the BKB.

Finally, we demonstrate the nonlocal skin effect in the electric circuit. Figure 4(a) shows the measured (blue dots) and calculated (red dots) admittance spectrum under open boundary conditions, where the grey area indicates the range covered by the spectrum under periodic boundary conditions in both directions. Obviously, the spectral areas under the two boundary conditions are quite different, indicating the presence of skin effect. Figure 4(b) presents the measured eigenstate distribution under open boundary conditions, where a bipolar skin effect is observed in the $x$-direction, while no skin effect appears in the $y$-direction. This can be explained by the distinct skin effect in the $x$- and $y$-directional ribbon structures. Figures 4(c) and 4(d) show the evolution of the skin effect in the $x$- and $y$-directions with respect to $k_y$ and $k_x$, respectively. As shown in Fig. 4(c), the eigenstates with $k_y \in (0, \pi)$ and $k_y \in (\pi, 2\pi)$ are nonlocally related through the MSNS reflection symmetry, and thus, are localized at opposite boundaries. At $k_y = 0$ and $k_y = \pi$, Bloch-wave-like extended states emerge, acting as translational states. Figure 4(d) shows the evolution of eigenstates in the $y$-direction, where no skin effect appears with respect to $k_x$. Figures 4(e) and 4(f) display the measured distribution of eigenstates, which clearly exhibit the nonlocal skin effect and match the theoretical results well.

In summary, we have theoretically and experimentally investigated the exceptional topology in the BKB and its associated nonlocal skin effect. In our non-Hermitian model, the MSNS reflection symmetry reduces the BZ to the BKB, opening an avenue to explore exceptional topology in non-orientable BZs. We have experimentally confirmed that the sum of the topological charges of all EPs in the BKB belongs to $2\mathbb{Z}$ rather than $0$ as in the conventional BZ. Moreover, due to the non-orientability of BKB, the topological charge of the EPs undergoes a phase transition when they cross antiparallel boundaries of the BKB, which has not been previously observed. Finally, the MSNS reflection symmetry nonlocally maps the skin effect at $k_y$ to that on the opposite side at $k_y + \pi$, offering potential applications in momentum-space mode filtering [61-63].

*Note added*. Recently, we become aware of the photonic [64] and phononic [65] works focus on the EP topology under periodic boundary conditions in the BKB.

**Acknowledgments**

This work was supported by the National Key Research and Development Program of China (Nos. 2022YFA1404900, 2022YFA1404501, 2023YFB2804701), the National Natural Science Foundation of China (Nos. 123B2069, 12304486, 12192253,

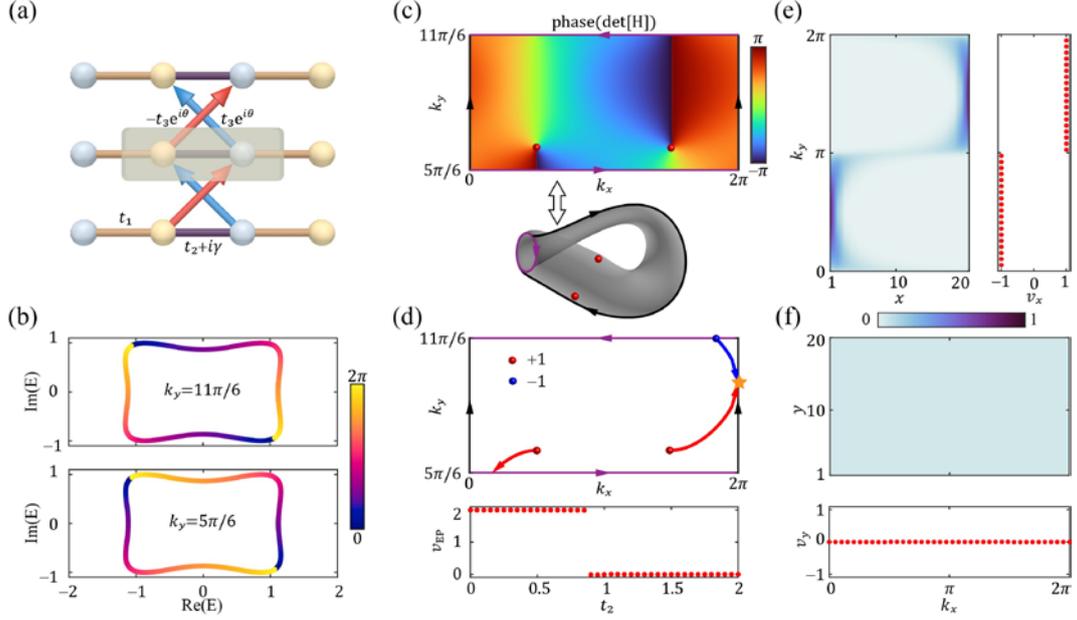

FIG. 1. Evolution of EPs and nonlocal skin effect in non-Hermitian Klein bottle model. (a) Schematic of the tight-binding model with two sites per unit cell. The red (blue) tubes represent the negative (positive) couplings. The intracell coupling term $i\gamma$ introduces the designed loss. (b) Evolution of bulk band dispersions at $k_y = 5\pi/6$ and $k_y = 11\pi/6$ with respect to $k_x$. Under the MSNS reflection symmetry, the two band dispersions wind in opposite directions. (c) Phase of $\det[H]$ in the BKB. Under MSNS reflection symmetry, the BZ can be reduced to half of its original size and forms a Klein bottle. There exist two EPs in the BKB, which correspond to vortices in the phase diagram. (d) Evolution of two EPs in the BKB as a function of $t_2$. When an EP crosses the antiparallel boundary of the BKB, its topological charge flips sign. The lower panel shows the evolution of the total topological charge $\nu_{\text{EP}}$ in the BKB as a function of $t_2$. (e) Evolution of the eigenstates and spectral winding number in the $x$-direction as a function of $k_y$. (f) Evolution of eigenstates and spectral winding number in the $y$-direction as a function of $k_x$. The parameters used are $t_1 = -1$, $\gamma = 1$, $t_3 = -0.5$, $\theta = 0$, and $t_2 = 0$.

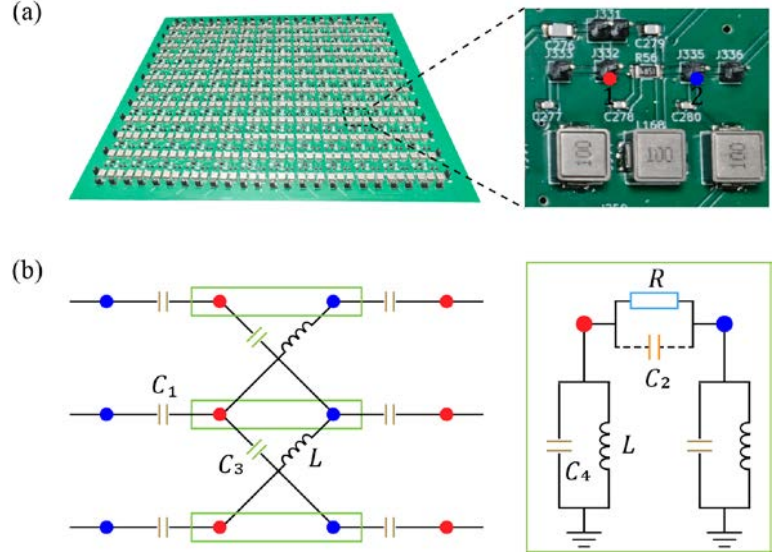

FIG. 2. Realization of the model in an electric circuit. (a) Photos of the electric circuit sample. The enlarged view shows the unit cell of the circuit board. (b) Schematic of the designed circuit, where the intracell couplings are highlighted in the green box. The capacitor $C_2$ can be controlled to determine whether it is connected to the circuit. Each node is grounded through an $LC$ resonator. The parameters of the circuit elements are chosen as $C_1 = 20 \text{ nF}$, $C_2 = 18 \text{ nF}$, $C_3 = 10 \text{ nF}$, $C_4 = 10 \text{ nF}$, $L = 9.5 \text{ μH}$, and $R = 15.4 \text{ Ω}$.

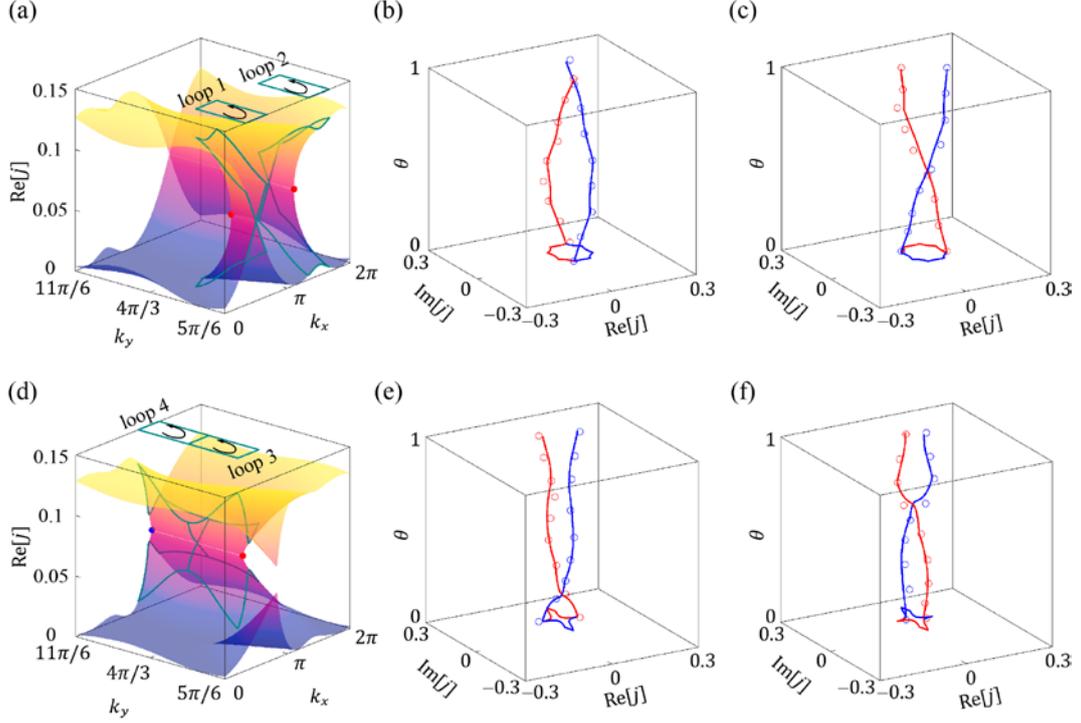

FIG. 3. Exceptional topology in the BKB. (a) Real parts of the admittance spectrum when $C_2$ is not connected, exhibiting two EPs with the same topological charge in the BKB. (b), (c) Measured stroboscopic evolution of eigenvalues along loop 1 (b) and loop 2 (c). The markers and lines represent the experimental and theoretical results, respectively. (d) Real parts of the admittance spectrum with $C_2$ connected. One of the EPs crosses the antiparallel boundary of BKB and undergoes a sign flip of its topological charge. (e), (f) Measured stroboscopic evolution of eigenvalues along loop 3 (e) and loop 4 (f). In (e) and (f), two bands wind in opposite directions, corresponding to the two EPs in (d) having opposite topological charges.

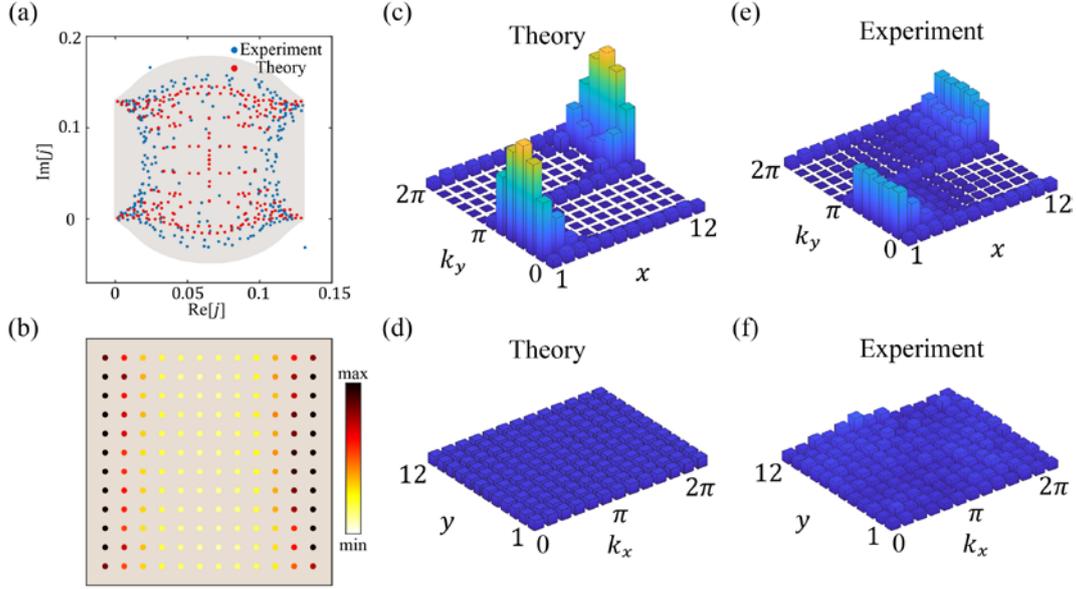

FIG. 4. Experimentally measuring the nonlocal skin effect. (a) Measured (blue dots) and calculated (red dots) admittance spectra under open boundary conditions. The grey area indicates the range covered by the spectrum under periodic boundary conditions in both directions. (b) Measured nonlocal skin effect under open boundary conditions. (c), (d) Calculated evolution of the skin effect in the $x$- and $y$-directions as a function of $k_y$ and $k_x$, respectively. (e), (f) Measured evolution of the skin effect in the $x$- and $y$-directions as a function of $k_y$ and $k_x$, respectively.